\documentclass[final,5p,times,twocolumn]{elsarticle}
\usepackage[T1]{fontenc}
\usepackage[utf8]{inputenc}
\usepackage{amsmath,amssymb,mathtools}
\usepackage{tikz}
\usepackage{microtype}
\usepackage[colorlinks=true,linkcolor=blue,citecolor=blue,urlcolor=blue]{hyperref}
\usetikzlibrary{arrows.meta}

\journal{Physics Letters B}

\newcommand{\dd}{\mathrm{d}}
\newcommand{\eps}{\epsilon}
\newcommand{\PV}{\operatorname{PV}}

\begin{document}

\begin{frontmatter}

\title{Characteristic Lightcone Sources in SO(1,3) Yang-Mills Theory}

\author[qmul]{Kaushlendra Kumar\corref{cor1}}
\ead{kaushlendra.kumar@qmul.ac.uk}
\cortext[cor1]{Corresponding author.}
\affiliation[qmul]{organization={School of Mathematical Sciences, Queen Mary University of London},
    addressline={Mile End Road},
    city={London},
    postcode={E1 4NS},
    country={United Kingdom}}

\begin{abstract}
The $\mathrm{SO}(1,3)$-symmetric reduction of Yang--Mills theory on Minkowski space yields a stress-energy tensor that is smooth on the timelike and spacelike Lorentz orbits but diverges on the lightcone that separates the two regions. We ask what consistent source this singularity represents. A natural cure is a real shift of the singular denominator. This does not regularize the source but instead moves the singular support onto a hyperboloid off the cone, where the corresponding source cannot remain both conserved and traceless. Our analysis shows that the displaced support cannot retain both properties, whereas a completion on the lightcone can. The matching condition fixes this completion up to one parameter $\chi$, which has a causal interpretation as the relative weight of the future and past cones. On regular constant-time slices, the completed source carries zero total four-momentum, so its physical content is a residual causal charge inferred from its curvature response.
\end{abstract}


\end{frontmatter}

\section{Introduction}
\noindent
Classical Yang--Mills theory on Minkowski space has limited explicit, highly symmetric time-dependent families. These include early $\mathrm{SO}(4)$-symmetric constructions \cite{DeAlfaroFubiniFurlan,Luscher}, shell-like configurations in conformally compactified Minkowski space \cite{GibbonsSteif}, and more recent families obtained from conformal reductions on de Sitter and anti-de Sitter spaces \cite{ILP17dS,ILP17AdS,LechtenfeldUnal,CosmicYM,AdSGaugeFields}. The present work builds on the explicit $\mathrm{SO}(1,3)$ construction of \cite{SO13PLB} and studies the source-theoretic status of its lightcone singularity. We use Minkowski coordinates $x^\mu=(t,\mathbf{x})$, $r=|\mathbf{x}|$, metric $\eta=\operatorname{diag}(-,+,+,+)$, and
\begin{equation}
s:=x\cdot x=\eta_{\mu\nu}x^\mu x^\nu=-t^2+r^2 .
\label{eq:s-def-intro}
\end{equation}
Away from the lightcone, $\mathrm{SO}(1,3)$ acts with three-dimensional orbits that foliate spacetime into homogeneous sheets: the timelike interior into hyperbolic spaces $H^3\simeq \mathrm{SO}(1,3)/\mathrm{SO}(3)$ and the spacelike exterior into de Sitter spaces $\mathrm{dS}_3\simeq \mathrm{SO}(1,3)/\mathrm{SO}(1,2)$. On each orbit the invariant gauge field reduces to a single profile $\phi(\ell)$ in the orbit-radial variable
\begin{equation}
\ell=\frac12\log |s|,
\label{eq:ell-def-intro}
\end{equation}
called $u$ in \cite{SO13PLB}; here the notation keeps $u,v$ free for standard lightcone coordinates below. The Yang--Mills equations then collapse to one-dimensional mechanics. The exceptional locus is the lightcone $s=0$, where the two foliations meet. There the orbit degenerates to the non-reductive coset $\mathrm{SO}(1,3)/\mathrm{ISO}(2)$, and the strictly invariant field is pure gauge. The discussion here is entirely classical and does not address unitarity of the non-compact gauge group inherited from this explicit sector.

 The physical content of the reduction is its stress--energy tensor. All dependence on the profile enters through one conserved number, the reduced mechanical energy
\begin{equation}
    \eps=\frac12\dot\phi^{\,2}-\frac12(\phi^2-1)^2,
    \qquad
    \dot\phi:=\frac{\dd\phi}{\dd\ell},
    \label{eq:eps-def-intro}
\end{equation}
in the normalization of \cite{SO13PLB}. On both sides of the lightcone, the gauge-invariant stress tensor reduces to the single covariant form
\begin{equation}
    T_{\mu\nu}
    =
    C\,\frac{4x_\mu x_\nu-\eta_{\mu\nu}\,s}{s^{3}},
    \qquad
    C:=\frac{\eps}{g^2},
    \label{eq:YM-tensor}
\end{equation}
where $g$ is the Yang--Mills coupling. The tensor is regular throughout the interior and exterior but diverges like $s^{-3}$ on the lightcone, where the reduction breaks down. The kink branch has $\eps=0$, so the stress tensor \eqref{eq:YM-tensor} and the associated lightcone defect vanish. This makes the present stress-energy problem source-trivial, so we restrict to the generic $\eps\neq0$ sectors. Both the tensor \eqref{eq:YM-tensor} and its superpotential were derived explicitly in \cite{SO13PLB}, where it was identified as a pure \emph{improvement} term. In terms of the scalar generator $\sigma_0$,
\begin{align}
    T_{\mu\nu}&=\bigl(\eta_{\mu\nu}\Box-\partial_\mu\partial_\nu\bigr)\sigma_0
    =\partial^\rho S_{\rho\mu\nu},
    \nonumber\\
    \sigma_0&=-\frac{C}{2s},
    \qquad
    S_{\rho\mu\nu} = C\,\frac{x_\rho\eta_{\mu\nu}-x_\mu\eta_{\rho\nu}}{s^2},
    \label{eq:S-improvement-intro}
\end{align}
so its energy and momentum densities are total spatial derivatives. Their integrals reduce to surface terms at spatial infinity, which vanish for the falloff $\sigma_0=O(r^{-2})$ (see Sec.~\ref{sec:obs} for details). The regularization proposed in \cite{SO13PLB} acts on this superpotential, $s \to s+\Delta$ in $S_{\rho\mu\nu}$ via real shift $\Delta$ (denoted $\delta$ there but we reserve this for Dirac distribution). The shifted superpotential gives the conserved completion
\begin{align}
    T^{\mathrm{reg}}_{\mu\nu}&=\partial^\rho S^{\mathrm{reg}}_{\rho\mu\nu}
    =C\,\frac{4x_\mu x_\nu-\eta_{\mu\nu}s+3\Delta\,\eta_{\mu\nu}}{(s+\Delta)^3},
    \nonumber\\
    S^{\mathrm{reg}}_{\rho\mu\nu}&=C\,\frac{x_\rho\eta_{\mu\nu}-x_\mu\eta_{\rho\nu}}{(s+\Delta)^2},
    \label{eq:Treg-intro}
\end{align}
which does not remove the singularity. It rather moves it to the hyperboloid
\begin{equation}
    \Sigma_\Delta:=\{x\in\mathbb{R}^{1,3}:s(x)=-\Delta\},
    \label{eq:SigmaDelta-intro}
\end{equation}
which is non-null for every real $\Delta\neq0$. Thus $T^{\mathrm{reg}}_{\mu\nu}$ is conserved but no longer traceless. We demonstrate in Section~\ref{sec:finite} that no real shift keeps both properties at once.

A consistent completion is instead supported on the lightcone. To see this, let $\delta_\pm(s):=\Theta(\pm t)\delta(s)$ be the future and past cone distributions and define
\begin{align}
    \sigma_\chi
    &=
    \sigma_{\mathrm{even}}+\chi\,\sigma_{\mathrm{odd}},
    \label{eq:sigmachi-intro}
    \\
    \sigma_{\mathrm{even}}
    &=
    -\tfrac{C}{2}\,\PV\!\Big(\tfrac1s\Big)+\tfrac{\pi C}{2}\,\delta(s),
    \nonumber\\
    \sigma_{\mathrm{odd}}
    &=
    \tfrac{\pi C}{2}\big(\delta_+(s)-\delta_-(s)\big),
    \nonumber
\end{align}
so that the resulting stress-energy tensor can be written as 
\begin{align}
T_{\mu\nu}=(\eta_{\mu\nu}\Box-\partial_\mu\partial_\nu)\sigma_\chi,\quad \Box=-\partial_t^2+\nabla^2.
\label{eq:completedSE}
\end{align}
This is a pure Callan--Coleman--Jackiw (CCJ) improvement source \cite{CCJ}, for which conservation is automatic and tracelessness is the harmonicity condition $\Box\sigma_\chi=0$. Requiring $\sigma_\chi$ to be harmonic through the origin fixes the total cone strength and the time-even core, leaving only the time-odd future/past weight $\chi$, with $\chi=+1,-1,0$ for retarded, advanced and time-symmetric completions. Section~\ref{sec:null} carries out this matching and shows that the real shift $\Delta$ serves only as a diagnostic and does not supply the required null-cone data. Within this construction, the completed source is determined up to $\chi$. The lesson is structural: for a distributional source it is the support class, not the tensor numerator, that decides whether conservation and tracelessness can coexist. The surviving datum $\chi$ is moreover physical rather than a regularization artifact. It is invisible to the real shift, yet it enters the gauge-invariant curvature pairing of Section~\ref{sec:obs}. Section~\ref{sec:conclusion} presents the outlook, while Appendix~\hyperref[app:conormal]{A} contains the distributional details and the no-go proof.

\section{Finite displacement and non-null support}
\label{sec:finite}
\noindent
The finite-$\Delta$ regularization fails in two complementary ways that can be made transparent by a one-parameter family as follows. Besides the conserved completion $T^{\mathrm{reg}}_{\mu\nu}$ of \eqref{eq:Treg-intro}, the other natural carrier is the bare denominator shift of \cite{SO13PLB}, which displaces only the pole of \eqref{eq:YM-tensor},
\begin{equation}
    T^{\Delta}_{\mu\nu}=C\,\frac{4x_\mu x_\nu-\eta_{\mu\nu}s}{(s+\Delta)^3}.
    \label{eq:TDelta}
\end{equation}
Writing $A_{\mu\nu}:=4x_\mu x_\nu-\eta_{\mu\nu}s$, the identities
\begin{equation}
    \eta^{\mu\nu}A_{\mu\nu}=0,
    \qquad
    \partial^\mu A_{\mu\nu}=18x_\nu,
    \qquad
    A_{\mu\nu}x^\mu=3s\,x_\nu
\end{equation}
make the tradeoff explicit. In the one-parameter Lorentz-covariant family
\begin{equation}
    T^{(a)}_{\mu\nu}
    =
    C\,\frac{A_{\mu\nu}+a\Delta\,\eta_{\mu\nu}}{(s+\Delta)^3},
    \label{eq:Ta}
\end{equation}
one finds
\begin{equation}
    T^{(a)\,\mu}{}_\mu
    =
    \frac{4aC\Delta}{(s+\Delta)^3},
    \qquad
    \partial^\mu T^{(a)}_{\mu\nu}
    =
    \frac{(18-6a)C\Delta\,x_\nu}{(s+\Delta)^4}.
    \label{eq:finite-a}
\end{equation}
Tracelessness fixes $a=0$ and conservation $a=3$, so for $\Delta\neq0$ the two requirements select different tensors: the bare shift $T^\Delta_{\mu\nu}$ ($a=0$) is traceless but not conserved, while the conserved completion $T^{\mathrm{reg}}_{\mu\nu}$ ($a=3$) is traceful. Their difference is $T^{\mathrm{reg}}_{\mu\nu}-T^\Delta_{\mu\nu}=\frac{3C\Delta\,\eta_{\mu\nu}}{(s+\Delta)^3}$, which vanishes pointwise for every fixed $s\neq0$ as $\Delta\to0$.

Both failures are governed by a single scalar, which the CCJ improvement structure makes explicit:
\begin{align}
    T_{\mu\nu}&=\mathcal I_{\mu\nu}\sigma =
    \partial^\rho U_{\rho\mu\nu}[\sigma],
    &
    \mathcal I_{\mu\nu}&:=\eta_{\mu\nu}\Box-\partial_\mu\partial_\nu,
    \nonumber\\
    U_{\rho\mu\nu}
    &:=
    \eta_{\mu\nu}\partial_\rho\sigma-\eta_{\rho\nu}\partial_\mu\sigma,
    &
    \eta^{\mu\nu}\mathcal I_{\mu\nu}\sigma&=3\Box\sigma ,
    \label{eq:improvement-superpotential}
\end{align}
where conservation is ensured by the antisymmetry $U_{\rho\mu\nu}=-U_{\mu\rho\nu}$ and tracelessness requires $\Box\sigma=0$. Away from the cone $T_{\mu\nu}=\mathcal I_{\mu\nu}\sigma_0$ with $\sigma_0=-\frac{C}{2s}$, and $U_{\rho\mu\nu}[\sigma_0]$ is the rank-three potential $S_{\rho\mu\nu}$ of \eqref{eq:S-improvement-intro}. The source problem is therefore to find a harmonic distributional extension of $\sigma_0$.

The real shift does not provide such an extension. It replaces the generator by $\sigma_\Delta=-\frac{C}{2(s+\Delta)}$, so that $T^{\mathrm{reg}}_{\mu\nu}=\mathcal I_{\mu\nu}\sigma_\Delta$, but $\sigma_\Delta$ fails to be harmonic:
\begin{equation}
    \Box\sigma_\Delta=\frac{4C\Delta}{(s+\Delta)^3}\neq0 .
    \label{eq:box-sigma-delta}
\end{equation}
The same non-vanishing scalar controls both failures:
\begin{equation*}
    T^{{\rm reg}\,\mu}{}_\mu=3\Box\sigma_\Delta,
    \qquad
    \partial^\mu T^\Delta_{\mu\nu}
    =-\frac34\,\partial_\nu\Box\sigma_\Delta .
\end{equation*}
It therefore determines the trace of $T^{\mathrm{reg}}_{\mu\nu}$ and, through its gradient, the non-conservation of $T^\Delta_{\mu\nu}$. To cure this problem, we need harmonic defect data, something the finite shift fails to provide.

This trace is itself a double divergence,
\begin{equation}
    T^{{\rm reg}\,\mu}{}_\mu=\partial_\mu\partial_\nu L^{\mu\nu},
    \quad
    L^{\mu\nu}=3\eta^{\mu\nu}\sigma_\Delta ,
    \label{eq:polchinski-L}
\end{equation}
which is precisely the condition for a conserved tensor to admit a traceless improvement $\Theta_{\mu\nu}=T^{\mathrm{reg}}_{\mu\nu}+\Delta T_{\mu\nu}$, with \cite[Eq.~(5)]{Polchinski}
\begin{align}
    \Delta T_{\mu\nu}
    &=\tfrac12\bigl(\partial_\mu\partial_\rho L^{\rho}{}_{\nu}+\partial_\nu\partial_\rho L^{\rho}{}_{\mu}-\Box L_{\mu\nu}-\eta_{\mu\nu}\partial_\rho\partial_\sigma L^{\rho\sigma}\bigr)
    \nonumber\\
    &\quad+\tfrac16\bigl(\eta_{\mu\nu}\Box L^{\rho}{}_{\rho}-\partial_\mu\partial_\nu L^{\rho}{}_{\rho}\bigr).
    \label{eq:polchinski-general}
\end{align}
For the pure-trace $L_{\mu\nu}=3\eta_{\mu\nu}\sigma_\Delta$ this reduces to $\Delta T_{\mu\nu}=-\mathcal I_{\mu\nu}\sigma_\Delta$, so the improvement is trivial,
\begin{equation}
    \Theta_{\mu\nu}
    =\mathcal I_{\mu\nu}\sigma_\Delta-\mathcal I_{\mu\nu}\sigma_\Delta
    =0 .
    \label{eq:Theta-zero}
\end{equation}
The conformal improvement therefore removes the whole finite-$\Delta$ tensor instead of completing it, and does not reproduce the nonzero source \eqref{eq:YM-tensor} as $\Delta\to0$. A nontrivial conserved and traceless completion that retains it still requires harmonic defect data, which only the characteristic null cone supplies.

\begin{figure}[t!]
\centering
\includegraphics[width=0.3\textwidth]{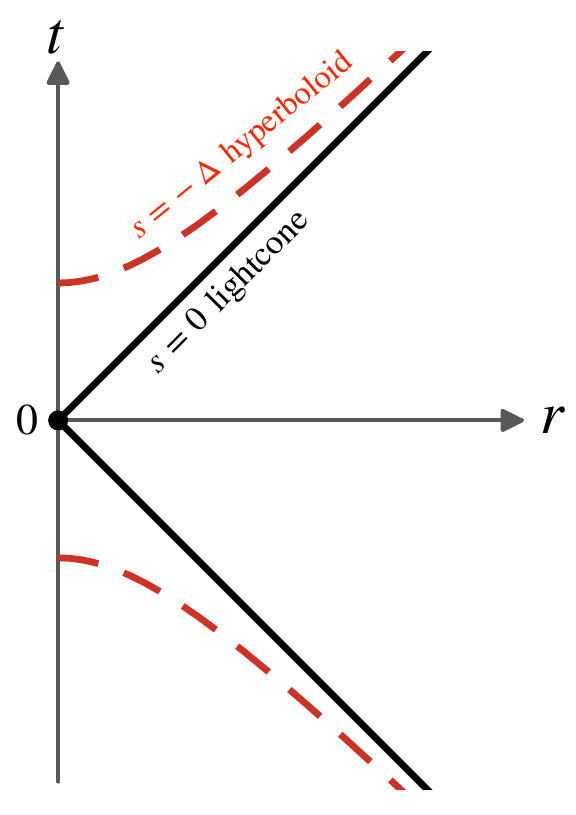}
\caption{Cross-sectional illustration of the shifted support. For $\Delta>0$, the singular carrier moves from the null cone $s=0$ (black) to the two-sheeted hyperboloid $s=-\Delta$ (red). The displaced support is non-null, hence non-characteristic for $\Box$, and approaches the cone only as $\Delta\to0^+$.}
\label{fig:support}
\end{figure}
Furthermore, as depicted geometrically in Figure~\ref{fig:support}, a real shift relocates the singular support rather than smoothing it. This is because every finite-$\Delta$ carrier is singular on the hyperboloid $\Sigma_\Delta$ \eqref{eq:SigmaDelta-intro}, whose conormal
\begin{equation}
    n_\mu=\partial_\mu s=2x_\mu,
    \qquad
    n^2=4s=-4\Delta,
\end{equation}
has nonzero length for every $\Delta\neq0$. This means that the principal symbol of $\Box$,
\begin{equation}
    p(\xi_\mu\dd x^\mu)=\eta^{\mu\nu}\xi_\mu\xi_\nu,
    \qquad
    p\bigl(\dd(s+\Delta)\bigr)=-4\Delta
    \quad\text{on}\quad
    \Sigma_\Delta,
    \label{eq:nonchar-finite}
\end{equation}
is nonzero and $\Sigma_\Delta$ is non-characteristic for $\Box$ \cite{Friedlander}. For $\Delta>0$ the shell splits into two sheets $\Sigma_{\Delta,\pm}:=\Sigma_\Delta\cap\{\pm t>0\}$, each a copy of $H^3$, whose ambient measures collapse to the future and past cone measures,
\begin{equation}
    \Theta(\pm t)\,\delta(s+\Delta)\,\dd^4x
    \;\longrightarrow\;
    \Theta(\pm t)\,\delta(s)\,\dd^4x
    =
    \tfrac12\rho\,\dd\rho\,\dd\Omega_2
\end{equation}
as $\Delta\to0^+$ (see Appendix~\hyperref[app:conormal]{A} for details), so at finite $\Delta$ the candidate defect lives on a non-null shell rather than on the cone. On this non-characteristic support any homogeneous distribution annihilated by $\Box$ must vanish, so the shell cannot carry a nontrivial traceless defect. This is a conormal obstruction, which we prove to all orders in Appendix~\hyperref[app:conormal]{A}. Among these carriers, the null cone is thereby singled out as the support on which $\Box$ admits the required nontrivial homogeneous solutions. We therefore turn our attention to this case in order to construct null-cone defect data in Section~\ref{sec:null}.

\section{Completion on the null cone}
\label{sec:null}
\noindent
On punctured Minkowski space $M^\times:=\mathbb{R}^{1,3}\setminus\{0\}$ the future and past cone distributions $\delta_\pm(s)=\Theta(\pm t)\delta(s)$ are well defined, since $\dd s\neq0$ except at the origin, and they satisfy
\begin{align}
    \partial_\mu\delta_\pm(s)&=2x_\mu\,\delta_\pm'(s), \label{eq:ident1}
    \\
    \partial_\mu\partial_\nu\delta_\pm(s)&=2\eta_{\mu\nu}\delta_\pm'(s)+4x_\mu x_\nu\delta_\pm''(s), \label{eq:ident2}
    \\
    \Box\,\delta_\pm(s)&=0
    \qquad\text{on }M^\times \label{eq:harmonicity},
\end{align}
as derived in Appendix~\hyperref[app:conormal]{A}. The last of these shows how the null cone supports nontrivial distributions annihilated by $\Box$ \cite{Friedlander}. Putting these back into the improvement operator $\mathcal I_{\mu\nu}$ of \eqref{eq:improvement-superpotential} produces a cone source,
\begin{equation}
    T^{\pm,\mathrm{cone}}_{\mu\nu}
    =\alpha_\pm\,\mathcal I_{\mu\nu}\delta_\pm(s)
    =-\alpha_\pm\bigl(2\eta_{\mu\nu}\delta_\pm'(s)+4x_\mu x_\nu\delta_\pm''(s)\bigr).
\end{equation}
It is conserved for any weight $\alpha_\pm$, by the same improvement identity, and its trace $3\alpha_\pm\Box\delta_\pm(s)$ also vanishes on $M^\times$. We therefore have an exact conserved and traceless source on $M^\times$, supported on the future or past cone.

On the full spacetime the origin must be restored. The bulk principal-value core carries a contact term while the cone defects carry the usual retarded/advanced fundamental solution identity in four dimensions \cite{Friedlander},
\begin{align}
    \Box\,\PV\!\left(\frac{1}{s}\right) &= -4\pi^2\delta^{(4)}(x), \label{eq:PV}
    \\
    \Box\,\delta_\pm(s) &= -2\pi\delta^{(4)}(x).
    \label{eq:Green}
\end{align}
Equivalently, the second identity identifies $-\delta_\pm(s)/(2\pi)$ as the retarded and advanced Green functions for the sign convention used here. In this way, we have a candidate ansatz for the null-cone completion:
\begin{equation}
    \sigma_{\alpha_+,\alpha_-}
    =
    -\frac{C}{2}\,\PV\!\Big(\frac1s\Big)
    +\alpha_+\delta_+(s)+\alpha_-\delta_-(s)
\end{equation}
for which one finds
\begin{align}
    \Box\,\sigma_{\alpha_+,\alpha_-}
    &=
    2\pi\bigl(\pi C-\alpha_+-\alpha_-\bigr)\delta^{(4)}(x),\\
     \alpha_++\alpha_-&=\pi C,
    \label{eq:matching}
\end{align}
where the matching law is forced by global harmonicity.

\medskip
\noindent
After matching, the one remaining freedom is how the fixed total strength $\alpha_++\alpha_-=\pi C$ is shared between the two cones. Writing
\begin{equation}
    \alpha_\pm=\frac{\pi C}{2}(1\pm\chi),
    \qquad
    \chi=\frac{\alpha_+-\alpha_-}{\alpha_++\alpha_-},
\end{equation}
makes the matching automatic and leaves $\chi$, the normalized future/past asymmetry, as the only parameter. This is the decomposition $\sigma_\chi=\sigma_{\mathrm{even}}+\chi\,\sigma_{\mathrm{odd}}$ of \eqref{eq:sigmachi-intro}, with the even and odd parts separately harmonic. The endpoints $\chi=\pm1$ place all the weight on the future or past cone, the retarded and advanced completions, while $\chi=0$ is the time-symmetric one. Requiring both weights $\alpha_\pm$ non-negative restricts $\chi$ to $[-1,1]$, and without that requirement $\chi$ is an arbitrary real parameter.

This completion cannot be reached by the real shift, as the limit $\Delta\to0$ on the regulated potential tends to the principal-value core alone,
\begin{equation}
    -\frac{C}{2}\PV\!\Big(\frac{1}{s+\Delta}\Big)
    \longrightarrow
    -\frac{C}{2}\PV\!\Big(\frac1s\Big),
\end{equation}
with no $\delta_\pm(s)$ cone term. The cone defect is therefore genuinely new data and is not a limiting case of the displaced shell. Such data would have to be supplied by a complex causal prescription or by a boundary condition on the cone, which lies outside the scope of the present work.

\section{Observable content of the completion}
\label{sec:obs}
\noindent
The natural next question is about the physical meaning of this completed stress-energy source. We show in the following that it carries no ordinary Poincar\'e charge, and its physical content rather lies in how it couples to external curvature. To that end, first notice that the pure-improvement form $T_{\mu\nu}=\mathcal I_{\mu\nu}\sigma_\chi$ in \eqref{eq:completedSE} implies that the translation-induced densities are spatial divergences on a constant-time slice,
\begin{align}
    T^{00}
    &=
    -\nabla^2\sigma_\chi
    =
    -\partial_i\partial_i\sigma_\chi,
    \nonumber\\
    T^{0i}
    &=
    \partial_t\partial_i\sigma_\chi
    =
    \partial_i\bigl(\partial_t\sigma_\chi\bigr).
\end{align}
Thus their spatial integrals reduce to surface terms at infinity,
\begin{align}
    P^0
    &=
    \int_t T^{00}\,\dd^3x
    =
    -\oint_{S^2_\infty}\nabla\sigma_\chi\cdot\dd\mathbf S,
    \nonumber\\
    P^i
    &=
    \int_t T^{0i}\,\dd^3x
    =
    \oint_{S^2_\infty}(\partial_t\sigma_\chi)\,\dd S_i.
    \label{eq:vanishing-P}
\end{align}
On any fixed slice $t\neq0$, the principal-value core has $\sigma_\chi=O(r^{-2})$, hence $\nabla\sigma_\chi=O(r^{-3})$ and $\partial_t\sigma_\chi=O(r^{-4})$, so both surface terms vanish. The cone-supported layers lie at the finite sphere $r=|t|$. Their spatial derivatives integrate to zero as distributions and do not create an additional internal boundary (see Appendix~\hyperref[app:conormal]{A} for cutoff argument). Therefore $P^\mu=0$ on every regular constant-time slice.

What it does carry is a curvature response, fixed by the same improvement structure. Let $g_{\mu\nu}=\eta_{\mu\nu}+h_{\mu\nu}$ be an external weak metric probe. The linear metric coupling and the scalar-curvature action producing the same weak probe response under standard metric variation are \cite{CCJ,BlaschkeEtAl}
\begin{align*}
    W[h]
    &:=
    \tfrac12\!\int T^{\mu\nu}h_{\mu\nu}\,\dd^4x,
    \\
    S_\sigma[g]
    &:=
    -\tfrac12\!\int\sqrt{-g}\,\sigma_\chi R[g]\,\dd^4x .
\end{align*}
For compactly supported $h_{\mu\nu}$, the scalar $R^{(1)}[h]$ is the linearized scalar curvature, appearing in $W[h]$ as a test function via integration by parts applied twice,
\begin{align}
    W[h]
    &=
    -\tfrac12\big\langle\sigma_\chi,R^{(1)}[h]\big\rangle,
    \nonumber\\
    R^{(1)}[h]
    &=
    \partial_\mu\partial_\nu h^{\mu\nu}-\Box h,
    \qquad h:=\eta^{\mu\nu}h_{\mu\nu},
\end{align}
where the pairing is defined in Appendix~\hyperref[app:conormal]{A}. The pairing provides a gauge-invariant readout of the fixed distribution $\sigma_\chi$, as $R^{(1)}[h]$ is invariant under compactly supported linearized diffeomorphisms of the probe.

An individual cone charge could be read from a one-sided probe after subtracting the known principal-value baseline fixed by $C$. The cleaner observable is the odd response, where this baseline cancels. We therefore use two time-reflected causal probes. Define $f_\pm:=R^{(1)}[h_\pm]$, with $f_+$ supported away from the origin in $t>0$ and $h_-(t,\mathbf x)=h_+(-t,\mathbf x)$, such that $f_-(t,\mathbf x)=f_+(-t,\mathbf x)$. We further normalize them as $\langle\delta_\pm(s),f_\pm\rangle=1$. For example, for an isotropic future probe, this reads $2\pi\int_0^\infty\rho f_+(\rho,\rho)\,\dd\rho=1$ by \eqref{eq:cone-pairing}. Since the principal-value core is time-even, its contribution is identical for the two probes and cancels in their response difference, such that, with $W_\pm:=W[h_\pm]$, we obtain
\begin{equation}
    W_+-W_-=-\frac12(\alpha_+-\alpha_-)
    =-\frac{\pi C}{2}\,\chi,
    \quad
    \chi=-\frac{2}{\pi C}(W_+-W_-).
    \label{eq:chi-response}
\end{equation}
The matching condition \eqref{eq:matching} then reconstructs the individual cone charges,
\begin{equation}
    Q_\pm=\alpha_\pm=\frac{\pi C}{2}(1\pm\chi),
    \qquad
    \chi=\frac{Q_+-Q_-}{Q_++Q_-}.
    \label{eq:chi-observable}
\end{equation}

\noindent\textit{Characteristic collar interpretation.}
The probe response in \eqref{eq:chi-response} measures the coefficients of cone-supported terms in $\sigma_\chi$. To identify what smooth local data can sit on a characteristic cone, we examine the harmonic equation on a collar of the cone, away from the tip. We use standard lightcone coordinates
\[
    u=t-r,\qquad v=t+r,\qquad s=-uv .
\]
The future cone is
\[
    u=0,\qquad v>0,
\]
while the past cone is $v=0$, $u<0$. A future collar means a neighbourhood of $u=0$ with $v$ bounded away from zero, so the cone tip is not part of this local calculation. Consider a local future-cone improvement layer of the form
\[
    \sigma_+=h(v,\Omega)\delta(u).
\]
Here $\Omega$ denotes the angular variables on $S^2$, and a prime below denotes $\partial_v$. In these coordinates,
\begin{equation*}
    \Box=-4\partial_u\partial_v
    +\frac{4}{v-u}(\partial_v-\partial_u)
    +\frac{4}{(v-u)^2}\Delta_{S^2},
\end{equation*}
and direct substitution gives
\begin{align*}
    \Box\bigl(h\delta(u)\bigr)
    &=
    \left(-4h'-\frac{4h}{v}\right)\delta'(u)
    \\
    &\quad+
    \left(\frac{4h}{v^2}+\frac{4h'}{v}
    +\frac{4\Delta_{S^2}h}{v^2}\right)\delta(u).
\end{align*}
The term $4h/v^2$ is a distributional product contribution. It follows from
\[
    a(u)\delta'(u)=a(0)\delta'(u)-a'(0)\delta(u),
    \qquad
    a(u)=-\frac{4h}{v-u}.
\]
For harmonicity on the collar, the independent coefficients of $\delta'(u)$ and $\delta(u)$ must vanish. This gives
\[
    h'+\frac{h}{v}=0,
    \qquad
    \frac{h}{v^2}+\frac{h'}{v}
    +\frac{\Delta_{S^2}h}{v^2}=0 .
\]
The first equation gives
\[
    h(v,\Omega)=\frac{q(\Omega)}{v},
\]
and the second then reduces to
\[
    \Delta_{S^2}q=0 .
\]
For smooth global angular data on $S^2$, only the constant spherical mode survives. Since
\[
    \delta_+(s)=\frac{1}{v}\delta(u)
\]
on the future collar, with $v^{-1}$ the root Jacobian, the only smooth future-cone layer is $\alpha_+\delta_+(s)$. The past collar gives the analogous layer $\alpha_-\delta_-(s)$. Thus $\chi$ is not a hidden angular shell profile. It is the normalized splitting of the fixed total cone strength between the two characteristic channels. A retarded no-incoming collar would select $\chi=+1$, the advanced counterpart would select $\chi=-1$, and a time-reversal-invariant collar would give $\chi=0$. Angular singularities or extra edge fields would define different sectors, not the smooth global collar considered here. A dynamical derivation of this boundary condition lies beyond the present letter.

The even/odd split has a clean meaning in terms of advanced/retarded Green functions, $\delta_\pm(s)=-2\pi\,G_{\mathrm{ret/adv}}$ \eqref{eq:Green}, in that their difference
\begin{equation}
    \sigma_{\mathrm{odd}}=-\pi^2 C\,\bigl(G_{\mathrm{ret}}-G_{\mathrm{adv}}\bigr)
    \label{eq:sigma-odd-PJ}
\end{equation}
is proportional to the Pauli--Jordan commutator function, a homogeneous solution of $\Box G=0$. The even core, on the other hand, combines the principal-value bulk with the time-symmetric retarded-plus-advanced cone contribution. A time-reversal-even probe gives $\langle\sigma_{\mathrm{odd}},R^{(1)}\rangle=0$, so the odd mode is invisible to time-symmetric measurements and can be detected only by causal probes that separate future from past.

\section{Conclusion and outlook}
\label{sec:conclusion}
\noindent
This work identifies the distributional source represented by the singular $\mathrm{SO}(1,3)$ Yang--Mills lightcone tensor. A real shift merely moves the carrier to the non-null hyperboloid $s=-\Delta$, where conservation and tracelessness select different tensors. The consistent completion remains on $s=0$ and is generated by the harmonic distribution $\sigma_\chi$ \eqref{eq:sigmachi-intro}. Matching fixes $Q_++Q_-=\pi C=\pi\eps/g^2$ and leaves only the normalized future/past asymmetry $\chi$.

The source has no integrated energy or momentum on regular constant-time slices, so $\chi$ is a causal cone charge rather than an energy. At linear order, time-reflected curvature probes isolate $\chi$, while time-symmetric probes are blind to it. Dirac's homogeneous radiative field provides the retarded-minus-advanced analogy \cite{Dirac1938}. Wheeler and Feynman instead begin with time-symmetric retarded-plus-advanced interactions and obtain an effective causal response from global absorber conditions \cite{WheelerFeynman}. These are analogies between causal Green functions, not identifications with electromagnetic radiation. The bulk quantity $C=\eps/g^2$ fixes the total cone strength $Q_++Q_-=\pi C$, while $\chi$ only partitions it as $Q_\pm=\frac{\pi C}{2}(1\pm\chi)$.

The main open problem is to derive the near-tip boundary condition dynamically by relaxing strict invariance near the cone and matching to the bulk Yang--Mills solution. A gravitational extension must also test the source against the admissibility conditions for distributional curvature \cite{GerochTraschen} and null-shell junction data \cite{BarrabesIsrael}. Since the present stress tensor contains derivatives of cone distributions, it is not an ordinary thin shell and this relation requires a separate analysis. Furthermore, under $x^\mu\to\lambda x^\mu$, the displacement scales as $\Delta\to\lambda^2\Delta$, so any fixed nonzero $\Delta$ introduces a length scale. By contrast, the matched null completion in \eqref{eq:completedSE} obeys $T_{\mu\nu}(\lambda x)=\lambda^{-4}T_{\mu\nu}(x)$, retaining the homogeneous scaling of the original four-dimensional Yang--Mills tensor. The broader conformal interpretation will be explored elsewhere.

\section*{Acknowledgements}
\noindent
The author thanks Olaf Lechtenfeld and Gabriel Pican\c{c}o Costa for past related collaborations and is grateful to the DFG for support through a Walter Benjamin Fellowship, project number 515782239.

\appendix

\section{Distributional shell calculus}
\label{app:conormal}
We collect the identities used in the main text. For a distribution $U$ and a smooth compactly supported test function $f$, the pairing $\langle U,f\rangle$ denotes the action of $U$ on $f$, with distributional derivative (from integration by parts) $\langle\partial_\mu U,f\rangle=-\langle U,\partial_\mu f\rangle$.

\medskip
\noindent\textbf{Cone and shell distributions.}
For a one-variable distribution $G$, its pullback $G(s)$ by $s(x)=-t^2+r^2$ is defined wherever $\dd s\neq0$: in adapted coordinates $(\zeta,y^a)$ with $\zeta=s$ and $\dd^4x=J(\zeta,y)\,\dd\zeta\,\dd^3y$,
\begin{equation}
    \langle G(s),f\rangle=\langle G(\zeta),h(\zeta)\rangle,
    \quad
    h(\zeta):=\int f(\zeta,y)\,J(\zeta,y)\,\dd^3y .
    \label{eq:pullback-pairing}
\end{equation}
In particular the root-splitting rule fixes the cone layers,
\begin{equation}
    \delta(g(t))=\sum_i\frac{\delta(t-t_i)}{|g'(t_i)|}
    \quad\Longrightarrow\quad
    \delta(s)=\frac{\delta(t-r)}{2r}+\frac{\delta(t+r)}{2r},
    \label{eq:delta-s-roots}
\end{equation}
so that $\delta_\pm(s)=\Theta(\pm t)\,\delta(s)$ are the future ($t=r$) and past ($t=-r$) cone distributions. Since $\dd s=2x_\mu\dd x^\mu$ vanishes only at the origin, the pullback is legitimate on $M^\times$, where the chain rule gives the distributional identities \eqref{eq:ident1} and \eqref{eq:ident2}. Together with $s\,\delta^{(n)}(s)=-n\,\delta^{(n-1)}(s)$, these identities give harmonicity $\Box\delta_\pm(s)=0$ \eqref{eq:harmonicity} on $M^\times$. The displaced shell is non-harmonic for all $\Delta\neq 0$,
\begin{equation}
    \Box\,\delta(s+\Delta)=8\delta'(s+\Delta)+4s\,\delta''(s+\Delta)=-4\Delta\,\delta''(s+\Delta).
    \label{eq:box-shifted-shell}
\end{equation}
The two meet in the limit: for $f$ supported away from the origin,
\begin{align}
    \big\langle\Theta(\pm t)\,\delta(s+\Delta),f\big\rangle
    &\xrightarrow{\;\Delta\to0^+\;}
    \big\langle\delta_\pm(s),f\big\rangle
    \nonumber\\
    &=
    \tfrac12\!\int_0^\infty\!\rho\,\dd\rho\,\dd\Omega_2\,
    f(\pm\rho,\rho\hat n).
    \label{eq:cone-pairing}
\end{align}
On the full space the cone tip carries the contact identities used in Section~\ref{sec:null}. The first of these \eqref{eq:PV} comes from Euclidean continuation of $\Delta_E(1/x_E^2)=-4\pi^2\delta^{(4)}$, while the second result \eqref{eq:Green} is the usual identification of $-\delta_\pm(s)/2\pi$ as the retarded and advanced fundamental solutions of $\Box$ \cite{Friedlander}.

\medskip
\noindent\textbf{Spatial integrals on regular slices.}
The charge calculation in Section~\ref{sec:obs} is understood on constant-time slices with $t\neq0$, for which the cone intersects $\mathbb R^3$ in the finite sphere $r=|t|$. Let $U_t$ denote a cone-supported distribution on such a slice and choose a smooth cutoff $\rho_R(\mathbf x)$ that equals one for $r\leq R$ and vanishes for $r\geq2R$. For sufficiently large $R$, the support of $\partial_i\rho_R$ is disjoint from the cone sphere, and therefore
\begin{equation}
    \big\langle\partial_i U_t,\rho_R\big\rangle
    =-\big\langle U_t,\partial_i\rho_R\big\rangle=0.
    \label{eq:cone-cutoff}
\end{equation}
The same argument applies after further spatial or time derivatives, and therefore the cone layers do not create an additional internal boundary term at $r=|t|$. The non-compact principal-value core is instead controlled by its falloff at spatial infinity. Moreover, at $t=0$ the cone tip does not define a regular pullback to the spatial slice. The four-momentum statement is therefore made on regular slices.

\medskip
\noindent\textbf{Local no-go.}
Let $P$ be a second-order operator with smooth coefficients and $\Sigma=\{\zeta=0\}$ a non-characteristic hypersurface. In adapted coordinates $(\zeta,y^a)$, only the leading-order term raises the transverse order,
\begin{align}
    P&=A(\zeta,y)\,\partial_\zeta^2+(\text{lower transverse order}),
    \nonumber\\
    A(0,y)&=p_2(x,\dd\zeta)\neq0 .
    \label{eq:nogo-operator}
\end{align}
A distribution on $\Sigma$ is locally a finite tower \cite[Thm.~2.3.5]{Hormander}, 
\begin{equation}
    U=\sum_{n=0}^N a_n(y)\,\delta^{(n)}(\zeta),
    \label{eq:shell-tower}
\end{equation}
and acting with \eqref{eq:nogo-operator}, the top layer of $PU$ becomes
\begin{equation}
    PU=A(0,y)\,a_N(y)\,\delta^{(N+2)}(\zeta)+(\text{lower transverse order}),
    \label{eq:nogo-top}
\end{equation}
which no lower-order or tangential term can reach. Since $A(0,y)\neq0$, $PU=0$ forces $a_N=0$, and descending induction gives $U=0$. For $P=\Box$ and $\zeta=s+\Delta$, the leading coefficient is
$A(0,y)=p_2(\dd\zeta)=-4\Delta\neq0$. Thus $\Box U=0$ on $\Sigma_{\Delta,\pm}$ forces $U=0$ for every finite $\Delta\neq0$. In other words, a nontrivial harmonic defect requires null support.

\end{document}